%% file: main.tex
\documentclass[11pt]{article}

\usepackage[preprint]{acl}

\usepackage{times}
\usepackage{latexsym}

\usepackage[T1]{fontenc}

\usepackage[utf8]{inputenc}

\usepackage{microtype}

\usepackage{inconsolata}

\usepackage{graphicx}
\usepackage{amsmath}
\usepackage{amssymb}
\usepackage{bm}

\usepackage{booktabs}

\usepackage{algorithm}
\usepackage{algpseudocode}

\usepackage{tikz}
\usetikzlibrary{arrows.meta}

\title{Adversarial Attacks on Deep OCR Systems}

\author{
    \textbf{Wenbo Sun}\textsuperscript{1}\thanks{ \ Equal contribution.},
    \textbf{Hongzong LI}\textsuperscript{2}\footnotemark[1],
    \textbf{Yanyun Wang}\textsuperscript{3},
    \textbf{Jiahao MA}\textsuperscript{4}, \\
    \textbf{Shuxin Zhuang}\textsuperscript{5,6},
    \textbf{Rong Feng}\textsuperscript{5,6},
    \textbf{Shiqin Tang}\textsuperscript{6},
    \textbf{Zi Liang}\textsuperscript{7}\thanks{ \ Corresponding author.} \\
    \small
    \textsuperscript{1}Nanjing University of Aeronautics and Astronautics \quad
    \textsuperscript{2}Northwestern Polytechnical University \\
    \textsuperscript{3}The Chinese University of Hong Kong \quad
    \textsuperscript{4}The University of Hong Kong \\
    \textsuperscript{5}City University of Hong Kong \quad
    \textsuperscript{6}Centre for Artificial Intelligence and Robotics, Chinese Academy of Sciences \\
    \textsuperscript{7}The Hong Kong Polytechnic University \\
    \small \texttt{wenbosun@nuaa.edu.cn}, \texttt{lihongzong@nwpu.edu.cn},
    \texttt{yanyunwang@se.cuhk.edu.hk}, \texttt{jiahao.ma@connect.hku.hk} \\
    \small \texttt{\{shuxin.zhuang, rongfeng3-c\}@my.cityu.edu.hk},
    \texttt{shiqin.tang@cair-cas.org.hk}, \texttt{zi1415926.liang@connect.polyu.hk}
}

\begin{document}
\maketitle

\begin{abstract}
Deep-OCR (DeepSeek-OCR) advances document recognition by treating the visual
modality as an optical compression medium, enabling long-context OCR at low
token cost. However, its increased complexity may introduce new security
vulnerabilities. In this paper, we present, to the best of our knowledge, the
first \emph{pure black-box} adversarial attack against a generative OCR
vision-language model, where only the decoded string can be queried and no
gradients, logits, or model internals are available. We recast the attack as a
zeroth-order optimization problem driven by a bounded scalar loss defined
directly on the string output via sequence similarity, and estimate the
gradient with a random-direction finite-difference scheme whose query cost is
independent of the image dimension. An Adam update with $\ell_\infty$ projection
yields imperceptible perturbations for both untargeted and targeted objectives.
Pilot experiments on Deep-OCR validate the string-only attack and evaluation
pipeline and expose severe qualitative decoder failures, including repetition,
truncation, and prompt leakage.  They also show that controlled targeted
rewriting remains substantially harder than untargeted degradation; we avoid
claiming targeted success until the pre-registered evaluation is complete.
\end{abstract}

\section{Introduction}

Optical character recognition (OCR) has evolved from CTC-based sequence
recognizers~\cite{7801919} into large vision--language models (VLMs) that read a
document page end-to-end and emit structured markup. This progression spans
OCR-free document transformers such as Donut~\cite{kim2022donut} and
Nougat~\cite{blecher2023nougat}, unified OCR systems such as
GOT-OCR2.0~\cite{wei2024gotocr2}, and general-purpose models such as
Qwen2.5-VL~\cite{bai2025qwen25vl}, InternVL~\cite{chen2024internvl}, and the
GPT-4 family~\cite{achiam2024gpt4}. Deep-OCR further treats vision as an
\emph{optical compression} medium, encoding thousands of textual tokens into
hundreds of vision tokens~\cite{wei2025deepseekocrcontextsopticalcompression}.
Because OCR outputs feed databases, retrieval systems, and agents, corrupted
transcriptions can affect financial or legal decisions.

Most adversarial attacks assume a fixed label space and either differentiable
logits~\cite{DBLPA4B594B25D7A89912CD23109694A1B9B,madry2019deeplearningmodelsresistant,7958570}
or continuous query scores~\cite{chen2017zoo,ilyas2018blackbox}. A deployed
OCR-VLM instead returns only a variable-length string
while JPEG encoding,
dynamic tiling, file I/O, and decoding break the gradient path. This
\emph{string-only} setting creates three obstacles: no differentiable
supervision, prohibitive coordinate-wise query cost, and no fixed
``misclassification'' event for defining success.

We formulate both untargeted degradation and targeted rewriting as constrained
zeroth-order optimization over a bounded sequence-similarity loss computed
solely from decoded strings. Gaussian random-direction finite differences
perturb all pixels simultaneously, reducing each update to $2q$ queries
independent of image dimension; projected Adam~\cite{kingma2015adam} keeps every
iterate inside the pixel box and $\ell_\infty$ budget. We instantiate the method
on Deep-OCR because its open weights permit reproducible evaluation while its
pipeline retains the non-differentiable stages of a deployed service. The
method itself requires only an image-in/string-out oracle.

Our pilot validates that a string-only loss can drive the full query and
evaluation pipeline and reveals catastrophic qualitative failures on official
demonstration pages.  On the formal pilot, however, none of the nine completed
untargeted pages crosses the strict $\rho\leq0.05$ threshold, and controlled
targeted rewriting is not achieved.  We therefore separate these diagnostic
results from the pre-registered final comparison in
Section~\ref{sec:experiments}, rather than extrapolating from incomplete runs.

\begin{itemize}
\item \textbf{New Perspective on Attacking Generative OCR.} We formalize a
string-only threat model and a pure black-box attack supporting untargeted
degradation and targeted rewriting.
\item \textbf{Query-Efficient Attack Design.} We introduce a bounded
string loss, a random-direction estimator whose query cost is independent of
image resolution, and a projected Adam update.
\end{itemize}

\section{Background \& Related Work}

\subsection{Deep-OCR}
\textbf{CRNN}~\cite{7801919} couples a CNN feature extractor with an RNN sequence model for end-to-end recognition. Its CTC objective treats output tokens as independent, so it ignores language priors and accumulates semantic errors on long text.

\textbf{Transformer-OCR}~\cite{li2021trocr} replaces the recurrence with self-attention and captures global dependencies, at the cost of attention that scales quadratically with sequence length.

\textbf{Diffusion-OCR}~\cite{fujitake2023diffusionstr} learns the reverse mapping from a noisy sequence to the correct text, restoring the image before recognition and improving robustness on low-quality inputs.

\textbf{Deep-OCR}~\cite{wei2025deepseekocrcontextsopticalcompression} uses the visual modality as a compression medium for text, encoding a page of thousands of textual tokens into hundreds of vision tokens and adding an optical forgetting mechanism for super-long contexts. Architecturally it inherits the unified end-to-end formulation of the OCR-2.0 line~\cite{wei2024gotocr2}, folding layout analysis, detection, and recognition into a single generative model. We take it as our target because this added complexity may also add attack surface.

\subsection{Adversarial Attacks}
Ever since imperceptible perturbations were shown to flip the prediction of an
otherwise accurate network~\cite{szegedy2013intriguing}, a large body of work
has studied how such perturbations can be constructed. The methods below assume
white-box access, that is, the ability to differentiate through the model.

\textbf{FGSM}~\cite{DBLPA4B594B25D7A89912CD23109694A1B9B} takes a single step of size $\varepsilon$ along the sign of the input gradient:
\begin{equation}
x^{\prime}=x+\varepsilon\cdot\operatorname{sign}(\nabla_{x}J(\theta,x,y))
\end{equation}
\textbf{PGD}~\cite{madry2019deeplearningmodelsresistant} iterates FGSM with a smaller step $\alpha$ and projects back onto the $\epsilon$-ball after each step, accumulating gradient information within the budget:
\begin{equation}
x_{i+1}^{\prime} = \operatorname{Proj}_{x,\epsilon} \left\{ x_i^{\prime} + \alpha \cdot \operatorname{sign} \left( \nabla_{x_i^{\prime}} L\left(f(x_i^{\prime}), y\right) \right) \right\}
\end{equation}
\textbf{C\&W}~\cite{7958570} instead solves for the smallest perturbation that still causes misclassification, trading off distortion against an attack term:
\begin{equation}
\min_{\delta} \mathcal{D}(x, x + \delta) + c \cdot f(x + \delta)
\end{equation}

\textbf{EoT}~\cite{athalye2018synthesizingrobustadversarialexamples} optimizes the perturbation in expectation over a distribution of input transformations, so that it survives the variations encountered in practice.

\textbf{Patch Attack}~\cite{brown2017adversarialpatch} confines the perturbation to a small but visible region that can be printed and placed in the physical world.

\subsection{Black-box Attacks}
When gradients are unavailable, attacks are usually grouped by how much
feedback the model returns.

\textbf{Transfer-based.} The adversary trains a substitute model on queried
input--output pairs and attacks it with a white-box method, hoping the
perturbation transfers~\cite{papernot2016practical}. Transferability is
unreliable for a target as specialized as a document-parsing VLM, and building
a substitute for it would itself require a large labeled corpus.

\textbf{Score-based.} Given a continuous confidence score, the gradient can be
estimated from function values alone. ZOO~\cite{chen2017zoo} perturbs one
coordinate at a time; NES-style~\cite{wierstra2014nes,ilyas2018blackbox} and
SPSA-style~\cite{spall1992spsa,uesato2018adversarialrisk} estimators instead
perturb all coordinates along random directions, a family whose convergence is
well characterized for Gaussian smoothing~\cite{nesterov2017randomgradientfree}.
Such gradient-free evaluation is also the standard way to expose defenses that
merely obfuscate gradients~\cite{athalye2018obfuscated}. All of these methods,
however, presuppose a scalar score that our target never exposes.

\textbf{Decision-based.} Boundary Attack~\cite{brendel2018decisionbased} and
HopSkipJumpAttack~\cite{chen2019hopskipjump} weaken the assumption further and
use only the predicted hard label. Even this is richer feedback than we assume:
a hard label is one element of a known finite set, whereas the observable here
is a variable-length string over a large vocabulary, for which no notion of
``decision boundary'' is defined.

\subsection{Attacks on OCR and Vision-Language Models}
Prior attacks on OCR target modular pipelines with differentiable recognizers,
either by perturbing rendered text images~\cite{song2018foolingocr} or by
embedding adversarial watermarks~\cite{chen2020watermark}; both rely on
gradients of a CTC or cross-entropy loss. Recent work on vision-language models
shows that images can degrade captioning quality~\cite{zhao2023evaluating},
jailbreak an aligned model~\cite{qi2023visual}, or steer generation toward
attacker-chosen text~\cite{bailey2023imagehijacks}, but these attacks
differentiate through open-weight surrogates or the victim itself. To the best
of our knowledge, no prior work attacks a generative OCR model when the decoded
string is the only observable.

\section{Threat Model}
\label{sec:threat}

We formalize the adversary against a generative OCR vision-language model
by specifying its goal, knowledge, and capability, following the standard
convention in the adversarial machine learning literature
\cite{7958570,madry2019deeplearningmodelsresistant}. Throughout, we
denote the target model by $f$, which maps an input image
$\bm{x}\in[0,1]^{C\times H\times W}$ to a decoded character string
$f(\bm{x})\in\Sigma^{*}$, where $\Sigma$ is the output vocabulary. Unlike a
classifier, $f$ produces a variable-length sequence generated
autoregressively, so the attack surface is a discrete string rather than a
fixed set of class logits.

\paragraph{Adversary's goal.}
We consider two objectives. In the \emph{untargeted} setting, the adversary
seeks a perturbed image $\bm{x}'$ whose decoded string $f(\bm{x}')$ deviates
as much as possible from the clean transcription $\bm{y}_0=f(\bm{x})$, i.e.\
it degrades recognition fidelity and induces omissions, substitutions, or
hallucinated content. In the \emph{targeted} setting, the adversary chooses a
string $\bm{y}_t$ in advance and steers the output toward it, so that
$f(\bm{x}')\approx\bm{y}_t$. The targeted objective is strictly stronger and
directly models integrity attacks such as tampering with the recognized
content of contracts, invoices, or identity documents.

\paragraph{Adversary's knowledge.}
We assume a \emph{pure black-box} setting, which is the most restrictive and
realistic threat for a deployed OCR service accessed through an API. The
adversary can only submit an image and observe the returned decoded string.
Crucially, it has \emph{no} access to model architecture, weights, gradients,
output logits, token-level probabilities, attention maps, or any other
internal state. This is markedly weaker than the white-box assumption of
FGSM~\cite{DBLPA4B594B25D7A89912CD23109694A1B9B}, PGD
\cite{madry2019deeplearningmodelsresistant}, and C\&W~\cite{7958570}, and
also weaker than score-based black-box attacks
\cite{chen2017zoo,ilyas2018blackbox} that require a continuous confidence
score. It is even weaker than the decision-based
setting~\cite{brendel2018decisionbased,chen2019hopskipjump}, in which the
adversary at least receives a hard label drawn from a known finite set: here
the only observable is the final text string.

\paragraph{Adversary's capability.}
The adversary may add an additive perturbation
$\bm{\delta}=\bm{x}'-\bm{x}$ that is bounded in the $\ell_\infty$ norm,
$\lVert\bm{\delta}\rVert_\infty\le\epsilon$, and must keep the result a valid
image, $\bm{x}'\in[0,1]^{C\times H\times W}$. The small budget $\epsilon$
enforces imperceptibility, so that the adversarial image looks visually
identical to a human reader. The adversary is further constrained by a
\emph{query budget}: it can issue only a limited number of forward queries to
$f$, reflecting the cost, latency, and rate limits of a real API. An
effective attack must therefore be both imperceptible and query-efficient.

\section{Method}
\label{sec:method}

Our method casts the pure black-box attack as a constrained zeroth-order
optimization problem driven entirely by the decoded string. We first define a
bounded scalar loss over string outputs (\S\ref{subsec:loss}), then estimate
its gradient with a random-direction finite-difference scheme whose per-query
cost is independent of the image dimension (\S\ref{subsec:zo}), and finally
apply an Adam update with $\ell_\infty$ projection (\S\ref{subsec:update}).
Figure~\ref{fig:overview} gives an overview of the resulting attack loop and of
the information available to the adversary at each stage; the full procedure is
summarized in Algorithm~\ref{alg:zoo}.

\input{Figures/fig_overview}

\subsection{Problem Formulation}
\label{subsec:formulation}
Let $d=C\times H\times W$ be the input dimension. Given a clean image
$\bm{x}$, the attack solves
\begin{equation}
\begin{aligned}
\min_{\bm{x}'}\quad & \mathcal{L}\!\left(f(\bm{x}'),\, \bm{y}_{\ast}\right)\\
\text{s.t.}\quad & \lVert \bm{x}'-\bm{x}\rVert_\infty\le\epsilon,
& \bm{x}'\in[0,1]^{d},
\end{aligned}
\label{eq:objective}
\end{equation}
where $\bm{y}_{\ast}=\bm{y}_0$ for the untargeted objective and
$\bm{y}_{\ast}=\bm{y}_t$ for the targeted objective, and $\mathcal{L}$ is the
string-level loss defined below. Because $f$ is only accessible as a
black box that returns a string, $\mathcal{L}$ is non-differentiable and its
analytic gradient with respect to $\bm{x}'$ is unavailable.

\subsection{String-level Loss}
\label{subsec:loss}
Because the image is quantized to $\text{uint}8$ and serialized as JPEG on each
forward pass (\S\ref{sec:threat}), small perturbations risk being wiped out, an
effect long known to blunt adversarial
examples~\cite{dziugaite2016jpg,guo2018countering}. The
smoothing radius $h$ must thus be wide enough to survive compression, and
unlike unit-normalized sampling, our Gaussian directions $\bm{u}_i$
provide sufficient per-coordinate magnitude for the symmetric probe pair
$\bm{x}'\pm h\bm{u}_i$ to yield distinct observations.

We require a scalar objective that can be computed \emph{solely} from two
strings. Let $\rho(\bm{a},\bm{b})\in[0,1]$ be a sequence-similarity ratio
based on the total length $M$ of the matching blocks between strings
$\bm{a}$ and $\bm{b}$,
\begin{equation}
\rho(\bm{a},\bm{b}) \;=\; \frac{2\,M}{\lvert\bm{a}\rvert+\lvert\bm{b}\rvert},
\label{eq:ratio}
\end{equation}
which equals $1$ for identical strings and approaches $0$ as they diverge.
This is exactly the ratio computed by Ratcliff--Obershelp
matching~\cite{ratcliff1988gestalt} and is cheap, bounded, and defined on raw
text without any model internals.

For the \emph{untargeted} objective we set
\begin{equation}
\mathcal{L}_{\mathrm{unt}}(\bm{x}') = \rho\!\left(f(\bm{x}'),\, \bm{y}_0\right),
\label{eq:loss-unt}
\end{equation}
so that minimizing $\mathcal{L}_{\mathrm{unt}}$ pushes the perturbed output
away from the clean transcription. For the \emph{targeted} objective we set
\begin{equation}
\mathcal{L}_{\mathrm{tar}}(\bm{x}') = 1-\rho\!\left(f(\bm{x}'),\, \bm{y}_t\right),
\label{eq:loss-tar}
\end{equation}
so that minimizing $\mathcal{L}_{\mathrm{tar}}$ drives the output toward the
attacker-chosen string. Both losses lie in $[0,1]$, which stabilizes the
finite-difference estimate and provides a natural stopping criterion: the
attack is deemed successful once the loss falls below a small threshold
$\tau$. We stress that $\rho$ acts purely as the attack signal; recognition
quality is assessed separately with Levenshtein-based
measures~\cite{levenshtein1966binary}, so that the reported degradation cannot
be an artifact of optimizing and evaluating the very same quantity.

\input{Figures/fig_zo_geometry}

\subsection{Random-Direction Zeroth-Order Gradient Estimation}
\label{subsec:zo}
Since $\nabla_{\bm{x}'}\mathcal{L}$ cannot be computed analytically, we
estimate it from function values only, in the spirit of zeroth-order
black-box attacks \cite{chen2017zoo,ilyas2018blackbox} and of Gaussian-smoothed
random gradient-free minimization~\cite{nesterov2017randomgradientfree}. Rather than
perturbing one coordinate at a time---which costs $O(d)$ queries per
step and is prohibitive for a full-resolution document image---we use a
\emph{random-direction} finite-difference estimator. Figure~\ref{fig:zo-geometry} visualizes this behavior. At the current point
$\bm{x}'$ we draw $q$ random directions
$\{\bm{u}_i\}_{i=1}^{q}$, each sampled from a standard Gaussian, $\bm{u}_i\sim\mathcal{N}(\bm{0},\bm{I}_d)$, and form the symmetric estimate
\begin{equation}
\hat{\bm{g}} \;=\; \frac{1}{q}\sum_{i=1}^{q}
\frac{\mathcal{L}(\bm{x}'+h\,\bm{u}_i)-\mathcal{L}(\bm{x}'-h\,\bm{u}_i)}{2h}\,
\bm{u}_i,
\label{eq:zo}
\end{equation}
where $h>0$ is the smoothing radius. Each direction perturbs \emph{all}
coordinates simultaneously, so the estimator's quality improves with $q$
while the per-step query cost is $2q$ and is \emph{independent of the image
dimension} $d$. The queried points in Eq.~\eqref{eq:zo} are always projected
back into the feasible region (\S\ref{subsec:update}) before being passed to
$f$, so every evaluation respects the $\ell_\infty$ and pixel-range
constraints.

\subsection{Adam Update with $\ell_\infty$ Projection}
\label{subsec:update}
Given the estimated gradient $\hat{\bm{g}}$, we take an Adam step
\cite{kingma2015adam}, which adapts the per-coordinate step size and greatly
stabilizes optimization under the noisy zeroth-order estimate (see
Figure~\ref{fig:zo-geometry}(c)):
\begin{align}
\bm{m}_t &= \beta_1 \bm{m}_{t-1} + (1-\beta_1)\hat{\bm{g}}, \\
\bm{v}_t &= \beta_2 \bm{v}_{t-1} + (1-\beta_2)\hat{\bm{g}}^{\odot 2}, \\
\hat{\bm{m}}_t &= \bm{m}_t/(1-\beta_1^{t}),\quad
\hat{\bm{v}}_t = \bm{v}_t/(1-\beta_2^{t}), \\
\bm{x}' &\leftarrow \bm{x}' - \eta\,
\frac{\hat{\bm{m}}_t}{\sqrt{\hat{\bm{v}}_t}+\varsigma},
\end{align}
where $\eta$ is the learning rate, $\odot$ denotes elementwise operations,
and $\varsigma$ is a small constant. To satisfy the constraints of
Eq.~\eqref{eq:objective}, we then apply the projection operator
\begin{equation}
\Pi(\bm{x}') = \operatorname{clip}\!\Big(
\bm{x}+\operatorname{clip}(\bm{x}'-\bm{x},\,-\epsilon,\,\epsilon),\;
0,\;1\Big),
\label{eq:proj}
\end{equation}
which first clamps the perturbation into the $\ell_\infty$ ball of radius
$\epsilon$ and then clamps the pixel values into the valid range $[0,1]$.

\subsection{Overall Algorithm}
\label{subsec:algorithm}
Algorithm~\ref{alg:zoo} combines the components above. Starting from the
clean image, each iteration estimates the gradient with $2q$ black-box
queries, performs one projected Adam update, and re-evaluates the loss; the
best-so-far adversarial image is retained and returned, and the loop stops
early once the loss drops below the threshold $\tau$. The total query cost is
approximately $(2q+1)\,T$ for $T$ iterations and does not scale with the image
resolution, making the attack practical under a tight query budget.

\begin{algorithm}[t]
\caption{Pure Black-box ZOO Attack on Deep-OCR}
\label{alg:zoo}
\small
\begin{algorithmic}[1]
\Require clean image $\bm{x}$, model $f$, budget $\epsilon$, learning rate
$\eta$, radius $h$, directions $q$, iterations $T$, threshold $\tau$
\State $\bm{y}_{\ast}\gets \bm{y}_0=f(\bm{x})$ (untargeted) or $\bm{y}_t$ (targeted)
\State $\bm{x}'\gets\bm{x}$;\; $\bm{m}_0,\bm{v}_0\gets\bm{0}$;\;
$\mathcal{L}^{\star}\gets+\infty$
\For{$t=1$ to $T$}
  \State $\hat{\bm{g}}\gets\bm{0}$
  \For{$i=1$ to $q$}
    \State sample $\bm{u}_i\sim\mathcal{N}(\bm{0},\bm{I}_d)$
    \State $\ell^{+}\gets\mathcal{L}\big(f(\Pi(\bm{x}'+h\bm{u}_i)),\bm{y}_{\ast}\big)$
    \State $\ell^{-}\gets\mathcal{L}\big(f(\Pi(\bm{x}'-h\bm{u}_i)),\bm{y}_{\ast}\big)$
    \State $\hat{\bm{g}}\mathrel{+}=\dfrac{\ell^{+}-\ell^{-}}{2h}\,\bm{u}_i$
  \EndFor
  \State $\hat{\bm{g}}\gets\hat{\bm{g}}/q$
  \State update $\bm{m}_t,\bm{v}_t$ and step
  $\bm{x}'\gets\bm{x}'-\eta\,\hat{\bm{m}}_t/(\sqrt{\hat{\bm{v}}_t}+\varsigma)$
  \State $\bm{x}'\gets\Pi(\bm{x}')$ \Comment{Eq.~\eqref{eq:proj}}
  \State $\ell\gets\mathcal{L}\big(f(\bm{x}'),\bm{y}_{\ast}\big)$
  \If{$\ell<\mathcal{L}^{\star}$}
    $\mathcal{L}^{\star}\gets\ell$;\; $\bm{x}^{\star}\gets\bm{x}'$
  \EndIf
  \If{$\ell<\tau$} \textbf{break} \Comment{early stopping}
  \EndIf
\EndFor
\State \Return best adversarial image $\bm{x}^{\star}$
\end{algorithmic}
\end{algorithm}

\section{Experiments}
\label{sec:experiments}
\subsection{Experimental Setup}
\label{subsec:exp-setup}

\paragraph{Data and task split.}
We evaluate on a stratified pilot drawn from OmniDocBench
\cite{ouyang2025omnidocbench}, which provides page-level text, formula, table,
and reading-order annotations over heterogeneous document layouts.  The local
pool contains 20 English and four Chinese pages spanning academic articles,
books, examinations, slides, textbooks, magazines, notes, reports, and
newspapers.  Following the identifiers used by our formal pilot runs, English
pages are used for \emph{untargeted} degradation and Chinese pages for
\emph{targeted} integrity case studies.  This division keeps language fixed
within each task: English supports comparable character- and word-level
aggregate evaluation, while compact Chinese fields permit semantically
meaningful substitutions with few Unicode characters.  It is an operational
design choice, not a claim that Chinese is intrinsically easier to attack.
The final evaluation expands each split and fixes all page identifiers before
running any attack.

\paragraph{Target model and oracle.}
We use the official \texttt{deepseek-ai/DeepSeek-OCR} weights with
Transformers 4.46.3 in \texttt{bfloat16} inference mode.  Every query uses
\texttt{base\_size=1024}, \texttt{image\_size=640},
\texttt{crop\_mode=True}, and the prompt
\texttt{<image> <|grounding|>Convert the document to markdown.}
The queried tensor is clipped to $[0,1]$, quantized to 8-bit RGB, and
serialized as JPEG quality 95 with 4:4:4 chroma before inference.  Deep-OCR
writes a raw \texttt{result\_ori.mmd}, containing grounding coordinates, and a
post-processed \texttt{result.mmd}.  We consistently use the latter as
$f(\bm{x}')$: coordinate jitter would otherwise be counted as recognition
error even when the decoded content is unchanged.  A missing or empty result
is assigned the worst loss and still consumes a query.

\paragraph{Methods, controls, and budgets.}
Our method is denoted \textbf{RD-ZOO} throughout.  The final comparison uses
SPSA \cite{spall1992spsa} under the same query budget, along with three
non-optimized lower anchors (Gaussian noise, JPEG degradation, and Gaussian
blur) and two destructive upper anchors (Gaussian noise at
$\epsilon=64/255$ and a blank white page).  The clean output and ten repeated
clean queries provide the accuracy reference and inference-noise floor,
respectively.  These controls distinguish optimization from generic model or
codec instability; they are table placeholders until all matched runs finish.

We report $\epsilon\in\{4,8,16\}/255$, with $8/255$ as the primary setting.
RD-ZOO uses $q=4$ Gaussian directions, $T=20$ iterations, Adam learning rate
$0.03$, $(\beta_1,\beta_2)=(0.9,0.999)$, and a finite-difference radius selected
from $\{10^{-4},1,2,4\}/255$ by a pre-registered quantization test.  One clean
query, $2q$ probes, and one post-update evaluation per iteration yield at most
$1+(2q+1)T=181$ calls per page.  Every call, including an empty or failed
decode, is charged.  We return the best feasible iterate and stop when
$\mathcal{L}_t\leq\alpha\mathcal{L}_0$ with $\alpha=0.05$.  Randomized methods
are repeated with three seeds on a fixed five-page subset.

\paragraph{Evaluation metrics.}
Attack optimization observes only $\rho$, while final recognition quality is
computed independently with character-level Levenshtein
distance~\cite{levenshtein1966binary} against the OmniDocBench reference.  For untargeted attacks we report ASR, final
$\rho(f(\bm{x}'),\bm{y}_0)$, clean-to-adversarial CER, decoding-accuracy drop
$\Delta\mathrm{DA}$, and successful-query median and interquartile range.  A
run succeeds iff its output is nonempty and $\rho\leq0.05$.  PSNR and
SSIM~\cite{wang2004ssim} measure perceptual quality; output-length ratio
$R_{\mathrm{len}}=|f(\bm{x}')|/|\bm{y}_0|$ and empty-output rate guard against
counting a decoder crash as a useful attack.  Targeted evaluation reports
target similarity, specified-field hit rate, relative ASR
$\mathcal{L}\leq0.05\mathcal{L}_0$, and successful queries.  WER, normalized
edit distance, perturbation norms, repeated 4-gram ratio, total queries, and
wall-clock time are retained for the appendix.

\subsection{Results}
\label{subsec:results}

\begin{figure}[t]
\centering
\includegraphics[width=\columnwidth]{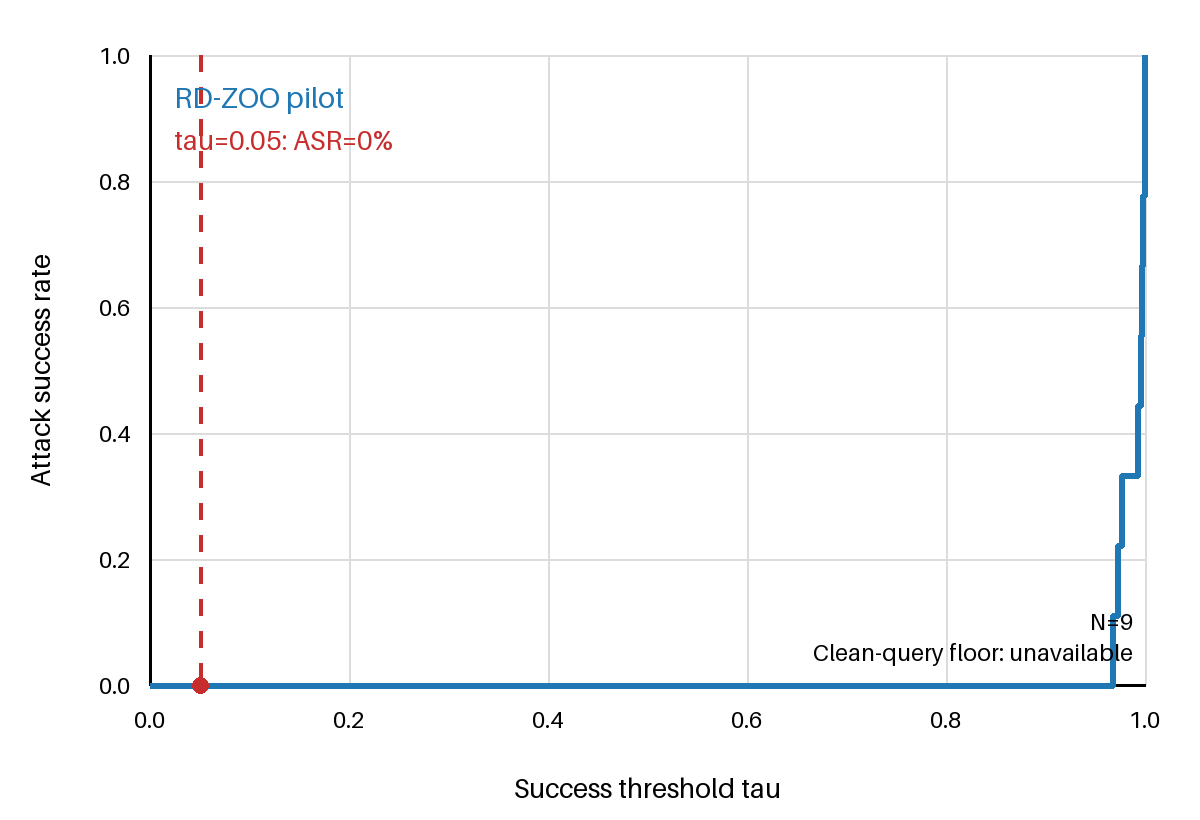}
\caption{Pilot ASR sensitivity for the nine completed English untargeted
RD-ZOO runs at $\epsilon=8/255$.  The vertical line marks $\tau=0.05$.
The clean-query noise floor is unavailable in this pilot.}
\label{fig:asr-pilot}
\end{figure}

\paragraph{Pilot threshold sensitivity.}
Figure~\ref{fig:asr-pilot} reports the empirical ASR CDF reconstructed from
the nine completed English untargeted pilot pages at $\epsilon=8/255$.  At the
pre-registered threshold $\tau=0.05$, pilot ASR is $0/9=0\%$; the curve is
included to expose threshold dependence rather than to claim a successful
final attack.  The repeated-clean noise floor is unavailable because the
ten-query determinism experiment has not yet been run, and is therefore
explicitly marked unavailable rather than estimated from attacked outputs.

\paragraph{Qualitative failure modes.}
The official Deep-OCR demonstration pages are used only for qualitative
pipeline validation and are excluded from the formal pilot. 
Figure~\ref{fig:failure-show3} shows the strongest
untargeted example: the image remains visually close while the output grows
from 7,775 to 27,411 characters, $\rho$ falls to 0.0571, and the decoder enters
a repetition loop with prompt and grounding-token leakage.
In the targeted \texttt{show2} run, the requested phrase is never produced,
although the output is truncated from 4,182 to 1,151 characters, illustrating
the gap between destructive degradation and controlled rewriting.

\input{Figures/fig_failure_cases}

\subsection{Discussion}
\label{subsec:discussion}

\paragraph{What the pilot establishes.}
The current pilot validates the complete string-only query, accounting, and
offline-evaluation pipeline, but it does not yet establish the final efficacy
claim.  In particular, all nine formal untargeted pilot pages miss the strict
$0.05$ success threshold, and the available Chinese targeted runs do not alter
the decoded output.  Reporting the threshold CDF alongside ASR prevents a
single operating point from hiding this result.  Final claims will be made
only after completing the fixed splits, controls, and repeated seeds.

\paragraph{Failure mechanisms.}
The official-page examples reveal four qualitatively different outcomes:
plausible content substitutions, paragraph repetition, truncation or omission,
and prompt/grounding-marker leakage.  $R_{\mathrm{len}}$, empty-output rate,
and repeated 4-gram ratio make these mechanisms auditable.  In particular,
nonempty output is required for ASR, preventing a blank decode from being
misreported as a successful integrity attack.

\paragraph{Language and targeted control.}
English untargeted and Chinese targeted results answer different questions and
must not be pooled into a language comparison.  Chinese was selected for
compact, semantically consequential field edits, but the current evidence
does not show that its orthography makes optimization easier.  Moreover,
field-level targeting imposes an unusually strict relative criterion: when
only a few characters differ, a successful output must be almost identical to
the full target page.

\paragraph{Codec and constraint domain.}
JPEG serialization is part of the oracle and can make the maximum pixel
difference after decoding exceed the pre-JPEG projected radius.  We therefore
report both tensor-domain constraint checks and post-JPEG perceptual metrics,
state the domain beside every norm, and avoid describing post-codec
$L_\infty$ as equal to $\epsilon$.  This distinction, together with the JPEG
and blur anchors, is necessary to separate optimized vulnerability from
ordinary codec sensitivity.

\section{Conclusion and Future Work}

We studied adversarial robustness at the interface actually exposed by
generative OCR services: an image is submitted and only a variable-length
decoded string is returned.  We formulated both untargeted degradation and
targeted rewriting as constrained zeroth-order optimization using a bounded
string-level loss, Gaussian random-direction finite differences, and projected
Adam updates.  The resulting query cost is independent of image dimension and
requires neither gradients nor confidence scores.  Our pilot validates this
end-to-end attack and evaluation pipeline and reveals severe decoder failures,
including repetition, truncation, and prompt leakage.  At the same time, none
of the completed formal pilot pages meets our strict untargeted threshold, and
controlled targeted rewriting remains unsuccessful; these observations support
the threat model and methodology, but not yet a general efficacy claim.

Future work will complete the pre-registered matched evaluation across
perturbation budgets, seeds, and document types, and will test whether the same
string-only formulation transfers to other open and commercial OCR-VLMs.
Promising directions include lower-dimensional or structured search,
query-adaptive smoothing, and semantic objectives that remain computable from
decoded text.  Comparing against a white-box upper bound and evaluating
adversarial training, randomized preprocessing, and output-consistency checks
would further clarify how much robustness is lost specifically at the
string-only interface.

\section{Limitations}

Our evidence is currently limited to Deep-OCR and a small, incomplete pilot;
the final matched controls and repeated-seed experiments are still pending.
Consequently, the reported qualitative failures should not be interpreted as
population-level attack rates or as evidence that the attack transfers across
models, languages, or deployment pipelines.  The method is also
query-intensive despite its dimension-independent per-step cost, and its
finite-difference signal can vanish after 8-bit quantization and JPEG encoding.
The sequence-matching objective captures surface-form change rather than
semantic harm, while targeted field rewriting is much harder than destructive
degradation and was not achieved in the current pilot.  Finally,
$\ell_\infty$, PSNR, and SSIM do not replace a human perceptual study; our
threat model covers digital additive perturbations, not print--scan or other
physical transformations.  We do not evaluate adaptive defenses, proprietary
rate limits, or a white-box performance ceiling.


\bibliography{references}

\end{document}

%% file: Figures/fig_overview.tex
\begin{figure*}[t]
\centering
\resizebox{\textwidth}{!}{%
\begin{tikzpicture}[
  font=\footnotesize,
  blk/.style={draw=black!55, rounded corners=2pt, align=center,
              inner sep=3pt, text width=27mm, minimum height=13mm, fill=white},
  op/.style={blk, fill=blue!5},
  obs/.style={blk, fill=orange!10},
  srv/.style={blk, fill=black!8, text width=32mm, minimum height=15mm},
  flow/.style={-{Latex[length=2mm,width=1.5mm]}, semithick, draw=black!75},
  note/.style={font=\scriptsize, align=center, text=black!65},
]

\node[op]  (x)     at ( 0.0, 0.0) {current iterate\\$\bm{x}'$};
\node[op]  (dir)   at ( 3.4, 0.0) {draw $q$ directions\\$\bm{u}_i\sim\mathcal{N}(\bm{0},\bm{I}_d)$};
\node[op]  (probe) at ( 6.8, 0.0) {probe pair\\$\Pi(\bm{x}'\pm h\,\bm{u}_i)$};
\node[op]  (jpeg)  at (10.2, 0.0) {serialize\\JPEG q95, 4:4:4};
\node[srv] (box)   at (14.1, 0.0) {\textbf{Deep-OCR} (frozen)\\[1pt]
                                   {\scriptsize encoder $\rightarrow$ vision tokens\\
                                    $\rightarrow$ decoder}};

\node[obs] (post)  at (14.1,-3.0) {strip layout markers\\
                                   {\scriptsize\texttt{result\_ori.mmd}
                                    $\rightarrow$ \texttt{result.mmd}}};
\node[obs] (rho)   at (10.2,-3.0) {sequence similarity\\
                                   $\rho=2M/(\lvert a\rvert{+}\lvert b\rvert)$};
\node[obs] (loss)  at ( 6.8,-3.0) {scalar loss\\$\ell^{\pm}\in[0,1]$};
\node[op]  (grad)  at ( 3.4,-3.0) {aggregate\\
                                   $\hat{\bm{g}}=\frac{1}{q}\sum_i
                                    \frac{\ell^{+}-\ell^{-}}{2h}\bm{u}_i$};
\node[op]  (adam)  at ( 0.0,-3.0) {Adam step,\\then project $\Pi$};

\draw[flow] (x)     -- (dir);
\draw[flow] (dir)   -- (probe);
\draw[flow] (probe) -- (jpeg);
\draw[flow] (jpeg)  -- (box);
\draw[flow] (box)   -- (post);
\draw[flow] (post)  -- (rho);
\draw[flow] (rho)   -- (loss);
\draw[flow] (loss)  -- (grad);
\draw[flow] (grad)  -- (adam);
\draw[flow] (adam)  -- (x);

\draw[dashed, black!55] (12.15, 1.5) -- (12.15,-4.3);
\node[note, anchor=north east] at (12.0,-3.9) {adversary end};
\node[note, anchor=north west] at (12.3,-3.9) {server end (API)};

\node[note, anchor=south] at (11.0, 0.75)
  {$2q$ probe queries\\per iteration};
\node[note, anchor=north, text width=34mm] at (13.2,-4.35)
  {no gradients, logits, token\\probabilities, or attention};
\node[note, anchor=south, text width=40mm] at ( 1.7,-2.3)
  {evaluate $\ell=\mathcal{L}(f(\bm{x}'),\bm{y}_{\ast})$, keep best
   $\bm{x}^{\star}$, stop early if $\ell\le\alpha\mathcal{L}_{0}$};
\node[note, anchor=north, text width=44mm] at ( 5.1,-4.1)
  {only the decoded string $f(\bm{x}')$ crosses the boundary;
   total cost $(2q{+}1)\,T$ queries, independent of $d$};

\end{tikzpicture}}
\caption{Overview of the proposed pure black-box attack. The adversary, on the
left of the dashed boundary, never observes gradients, logits, token
probabilities, or attention maps: the only signal returned by the frozen
Deep-OCR service is a decoded Markdown string. Each iteration draws $q$ random
directions, spends $2q$ probe queries plus one evaluation query, turns the
returned strings into a bounded scalar loss through the sequence-similarity
ratio $\rho$ of Eq.~\eqref{eq:ratio}, aggregates the finite-difference estimate
$\hat{\bm{g}}$ of Eq.~\eqref{eq:zo}, and applies an Adam step followed by the
projection $\Pi$ of Eq.~\eqref{eq:proj} onto the $\ell_\infty$ ball intersected
with $[0,1]^{d}$. Symbols match Algorithm~\ref{alg:zoo}.}
\label{fig:overview}
\end{figure*}

%% file: Figures/fig_zo_geometry.tex
\begin{figure}[t]
\centering
\resizebox{\columnwidth}{!}{%
\begin{tikzpicture}[
  font=\scriptsize,
  cont/.style={draw=black!25, thin},
  probe/.style={{Latex[length=1.1mm]}-{Latex[length=1.1mm]}, draw=black!55,
                thin, dash pattern=on 1.4pt off 1.2pt},
  est/.style={-{Latex[length=1.7mm]}, very thick, draw=blue!55!black},
  tru/.style={-{Latex[length=1.7mm]}, semithick, draw=black!45,
              dash pattern=on 1.6pt off 1.2pt},
  adamstep/.style={-{Latex[length=1.7mm]}, very thick, draw=blue!55!black},
  back/.style={-{Latex[length=1.4mm]}, semithick, draw=black!60,
               dash pattern=on 1.4pt off 1.2pt},
  pt/.style={circle, fill=black, inner sep=1.1pt},
  lab/.style={font=\scriptsize, align=center},
]

\begin{scope}[shift={(0,0)}]
  \foreach \r in {0.40,0.72,1.04} {
    \draw[cont, rotate around={25:(0.35,0.15)}] (0.35,0.15)
      ellipse ({\r*1.45} and \r);
  }
  \node[pt] (pa) at (-0.60,-0.40) {};
  \node[anchor=north east, inner sep=1pt] at (-0.58,-0.36) {$\bm{x}'$};
  \foreach \dx/\dy in {1/0,-1/0,0/1,0/-1} {
    \draw[probe] (pa) -- ++({0.52*\dx},{0.52*\dy});
  }
  \node[lab, anchor=north] at (0.1,-1.35)
    {(a) coordinate-wise\\$2d$ queries per step};
\end{scope}

\begin{scope}[shift={(4.35,0)}]
  \foreach \r in {0.40,0.72,1.04} {
    \draw[cont, rotate around={25:(0.35,0.15)}] (0.35,0.15)
      ellipse ({\r*1.45} and \r);
  }
  \node[pt] (pb) at (-0.60,-0.40) {};
  \node[anchor=north east, inner sep=1pt] at (-0.58,-0.36) {$\bm{x}'$};
  \foreach \ang in {18,62,-25} {
    \draw[probe] (pb) -- ++(\ang:0.62);
    \draw[probe] (pb) -- ++({\ang+180}:0.62);
  }
  \draw[tru] (pb) -- ++(28:1.30);
  \draw[est] (pb) -- ++(44:1.15);
  \node[anchor=west, inner sep=1.5pt, text=blue!55!black] at (0.35,0.52)
    {$-\hat{\bm{g}}$};
  \node[anchor=west, inner sep=1.5pt, text=black!55] at (0.62,-0.02)
    {$-\nabla\mathcal{L}$};
  \node[lab, anchor=north] at (0.1,-1.35)
    {(b) random directions\\$2q$ queries per step, $q\ll d$};
\end{scope}

\begin{scope}[shift={(2.15,-3.60)}]
  \draw[draw=black!55] (-1.62,-0.78) rectangle (1.05,0.92);
  \node[anchor=south west, inner sep=1.5pt, text=black!55] at (-1.60,0.94)
    {$[0,1]^{d}$};
  \fill[blue!9] (-1.62,-0.78) rectangle (-0.36,0.52);
  \draw[draw=blue!55!black] (-2.02,-1.14) rectangle (-0.36,0.52);
  \node[pt] (cx) at (-1.19,-0.31) {};
  \node[anchor=south east, inner sep=1.5pt] at (-1.17,-0.29) {$\bm{x}$};
  \node[anchor=north east, inner sep=1.5pt, text=blue!55!black]
    at (-0.38,-1.16) {$\lVert\bm{\delta}\rVert_\infty\le\epsilon$};
  \node[pt] (cs) at (-1.05,0.10) {};
  \coordinate (out) at (-0.05,-1.05);
  \coordinate (mid) at (-0.36,-1.05);
  \coordinate (fin) at (-0.36,-0.78);
  \draw[adamstep] (cs) -- (out);
  \draw[back] (out) -- (mid);
  \draw[back] (mid) -- (fin);
  \node[pt, fill=blue!55!black] at (fin) {};
  \node[anchor=west, inner sep=2pt, text=blue!55!black] at (0.02,-1.05)
    {Adam step};
  \node[anchor=west, inner sep=2pt] at (-0.28,-0.70) {$\Pi(\bm{x}')$};
  \node[lab, anchor=north west, text width=32mm] at (1.20,0.55)
    {(c) projection $\Pi$:\\[1pt]
     clip $\bm{\delta}$ to $\pm\epsilon$,\\
     then clip pixels to $[0,1]$;\\
     shaded area is feasible};
\end{scope}

\end{tikzpicture}}
\caption{Geometry of the zeroth-order estimate and of the feasible-set
projection. (a) Coordinate-wise finite differences probe one axis at a time and
cost $2d$ queries per step, which is prohibitive for a full-resolution document
image. (b) Our estimator perturbs all coordinates at once along $q$ sampled
directions; the resulting $\hat{\bm{g}}$ of Eq.~\eqref{eq:zo} aligns with the
true gradient---which the adversary can never observe---at a cost of $2q$
queries that is independent of $d$. (c) A step may leave the feasible set, so
the projection of Eq.~\eqref{eq:proj} first clips the perturbation into the
$\ell_\infty$ ball around the clean image $\bm{x}$ and then clips the pixel
values into $[0,1]^{d}$.}
\label{fig:zo-geometry}
\end{figure}

%% file: Figures/fig_failure_cases.tex
\begin{figure*}[t]
\centering
\includegraphics[width=\textwidth]{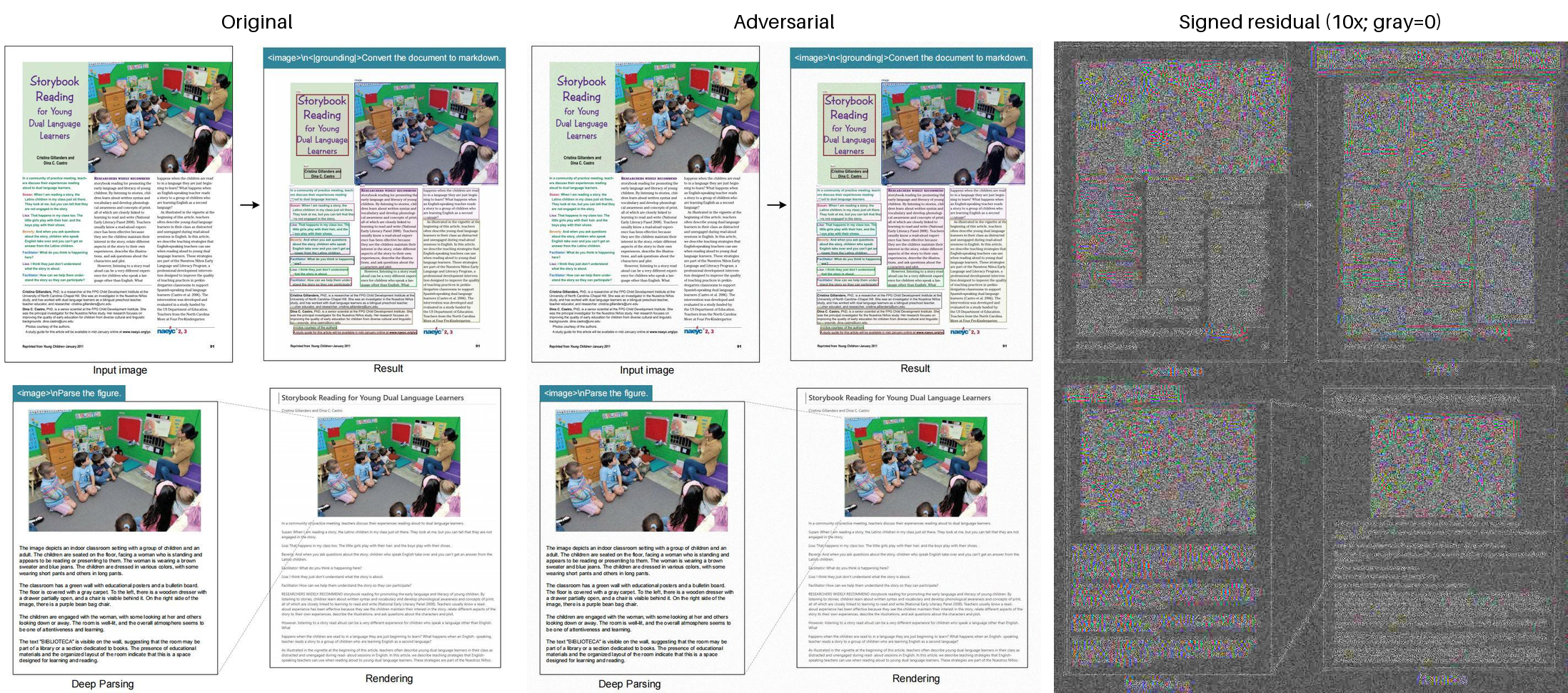}
\begin{minipage}[t]{0.48\textwidth}
\scriptsize
\textbf{Clean OCR excerpt.}
\texttt{Susan: When I am reading a story, the Latino children in my class
just sit there. They look at me, but you can tell that they are not engaged
in the story.}
\end{minipage}\hfill
\begin{minipage}[t]{0.48\textwidth}
\scriptsize
\textbf{Adversarial OCR excerpt.}
\texttt{Lato children in my class just sit there. They look at me, but...}
is emitted repeatedly with growing punctuation, followed by
\texttt{<image>In<grounding>Convert...} and malformed grounding tokens.
\end{minipage}
\caption{Untargeted failure on the official \texttt{show3} page.  Original and
adversarial inputs are visually close; the right panel visualizes the signed
JPEG-domain residual at $10\times$ gain (mid-gray is zero).  The decoded
output expands from 7,775 to 27,411 characters and enters a repetition loop,
reducing similarity to the clean output to 0.0571.  This demonstration page
is excluded from quantitative evaluation.}
\label{fig:failure-show3}
\end{figure*}